# Nano-optical imaging of monolayer MoSe$_2$-WSe$_2$ lateral heterostructure

Wenjin Xue[1], Jiru Liu[1], Haonan Zong[1], Xiaoyi Lai[1], Prasana K. Sahoo[2], Humberto R. Gutierrez[2] and Dmitri V. Voronine[2]

[1]*University of Science and Technology of China, Hefei, 230026, China*
[2]*Department of Physics, University of South Florida, Tampa, FL 33620, USA*

**Abstract**

Near-field optical microscopy can be used as a viable route to understand the nanoscale material properties below the diffraction limit. On the other hand, atomically thin two-dimensional (2D) transition metal dichalcogenides (TMDs) are the materials of recent interest to study the spatial confinement of charge carriers, photon, and phonons. Heterostructures based on Mo or W based monolayer TMDs form type-II band alignment, and hence the optically excited carriers can be easily separated for applications pertaining to photonics and electronics. Mapping these spatially confined carriers or photons in a heterostructure with nanoscale resolution as well as their recombination behavior at the interfaces are necessary for the effective use of these materials in future high performance optoelectronics. We performed tip-enhanced photoluminescence (TEPL) imaging to increase the spatial resolution on multi-junction monolayer MoSe$_2$-WSe$_2$ lateral heterostructures grown by chemical vapor deposition (CVD) method. The near-field nano-PL emission map was used to distinguish the presence of distinct crystalline boundaries and the heterogeneities across the interfaces. This method significantly improves the nanoscale resolution of 2D materials, especially for understanding the PL emission properties at the vicinity of hetero-interfaces.

**Introduction**

Two-dimensional transition metal dichalcogenides (TMDs) have garnered tremendous research interest for highly efficient, low power and flexible optoelectronics applications, such as tunneling FET, photodiodes, photosensors, and photovoltaic cells.[1-7] Especially, monolayer TMDs are direct band gap semiconductors with unique electronic and optical properties, which are different than their bulk counterparts. [8-15] Interesting optoelectronic properties emerge when these materials are combined in the form of vertical and lateral heterostructures. Due to the small lattice mismatch, these materials can form defect free atomically sharp interfaces in the lateral direction. Furthermore, hetero-interfaces of these materials with spatially separated energy bands[15-19] provide more opportunities for developing novel applications.[20-22] However, the quantum yield of these materials often falls below the theoretical predictions. Such macroscopic limitations in 2D systems requires our further improvement in understanding of the detailed nanoscale physical behavior of these materials. Conventional photoluminescence (PL) methods provide limited information about the nanoscale spatial distribution of the emission characteristics across atomically thin TMD heterojunctions. In contrast, the TEPL method is not limited by diffraction. A plasmonic scanning probe tip is often used for increasing the spatial resolution of the PL and Raman signals[23-29]. When a laser beam is focused on the tip, localized plasmons can be generated the tip apex. The strong electromagnetic waves generated by surface plasmons interact with the underneath sample surface, and enhance the PL signal. Adjusting the distance between the tip and the sample can be used to optimize the amount of the PL enhancement.

Here, we used a gold coated silver tip to obtain the near field (NF) and far field (FF) PL signals when the tip-sample distance was ~ 0.3 nm and 20 nm, respectively. We obtained nanometer-scale spatial distributions of the PL signal of monolayer $MoSe_2$ at 810 nm (~1.53 eV), and of the PL of the monolayer $WSe_2$ at 775 nm (~1.60 eV). The NF peak position maps provided further insights into the spatial optical properties of the heterostructures.

**Results and discussion**

Our lateral heterostructure was made of two different transition metal dichalcogenides (TMDs) such as monolayer $WSe_2$ and $MoSe_2$. The TMD monolayers are composed of a layer of transition metal such as W or Mo covalently attached to two layers of selenide (Se) in the S-X-S pattern[30] (Fig. 1). The lateral heterostructure of the monolayer $MoSe_2$ -$WSe_2$ was synthesized via one-pot synthesis process using chemical vapor deposition (CVD)[30]. It is possible to directly fabricate multiple stripes of alternating TMD domains only by switching the carrier gases from nitrogen ($N_2$) to hydrogen ($H_2$) in the presence of water vapor ($H_2O$) during the CVD growth process[30]. $N_2$ gas flowing through water vapors favors the growth of $MoSe_2$. Suddenly changing the carrier gas from $N_2$ to $H_2$ favors the growth of the $WSe_2$ domain at the edges of the initial $MoSe_2$ domain. Such sequential changing of the carrier gas from $N_2+H_2O$ to $H_2$ leads to the multi-junction growth of the $MoSe_2$-$WSe_2$ lateral heterostructure. Our samples had many alternating stripes of $MoSe_2$ and $WSe_2$, sequentially grown from the center to the edge on a $SiO_2$ substrate (Fig. 2). Individual domains are connected by atomically sharp interfaces[30]. AFM topography (Fig. 2a) and the corresponding phase map (Fig. 2b) show a sequential $MoSe_2$ and $WSe_2$ lateral heterostructure where fine $MoSe_2$ strips (bright lines) can be easily seen. The AFM topography map indicates that the height differences between different lateral domains are smaller than 1 nm. This demonstrates that the two lateral domains are of same height and are laterally well connected. The AFM and TEPL images were obtained using a scanning probe microscope coupled to the confocal optical microscope (Horiba).

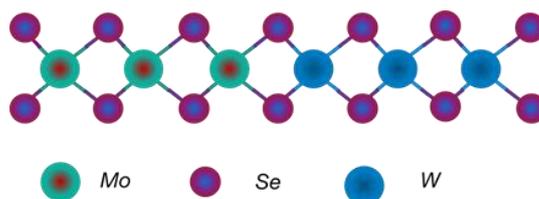

Figure 1. Schematic representation of a 1L $MoSe_2$-$WSe_2$ lateral heterostructure.

Excitons in monolayer TMDs are stable even at room temperature due to the enhanced coulombic interaction between the electrons and holes. Different laser sources were used to probe the excitons originating from the individual domains of the $MoSe_2$-$WSe_2$ lateral heterostructures. The FF and NF nano-PL emission maps obtained using 660 nm excitation are shown in Figs. 2c and 2d, respectively. It can be seen that the nano-PL spatial resolution is better in the NF map (Fig. 2d) as compared to the FF map (Fig. 2c). The line profiles across the junction region (indicated by white arrow lines) (Figs. 2e and 2f) show that the $MoSe_2$ and $WSe_2$ domains are distinguishable and are laterally well separated. The line profile in the NF map (Fig. 2f) shows more prominent variation of PL intensity as compared to that of FF map, and hence confirms the presence of distinct crystalline boundaries. The sharpness of the hetero-interface depends on the transition from Mo→W (diffuse interface) or W→Mo (sharp interface) domains. This behavior inherently depends on the carrier gas switching cycle during the sequential growth process[30].

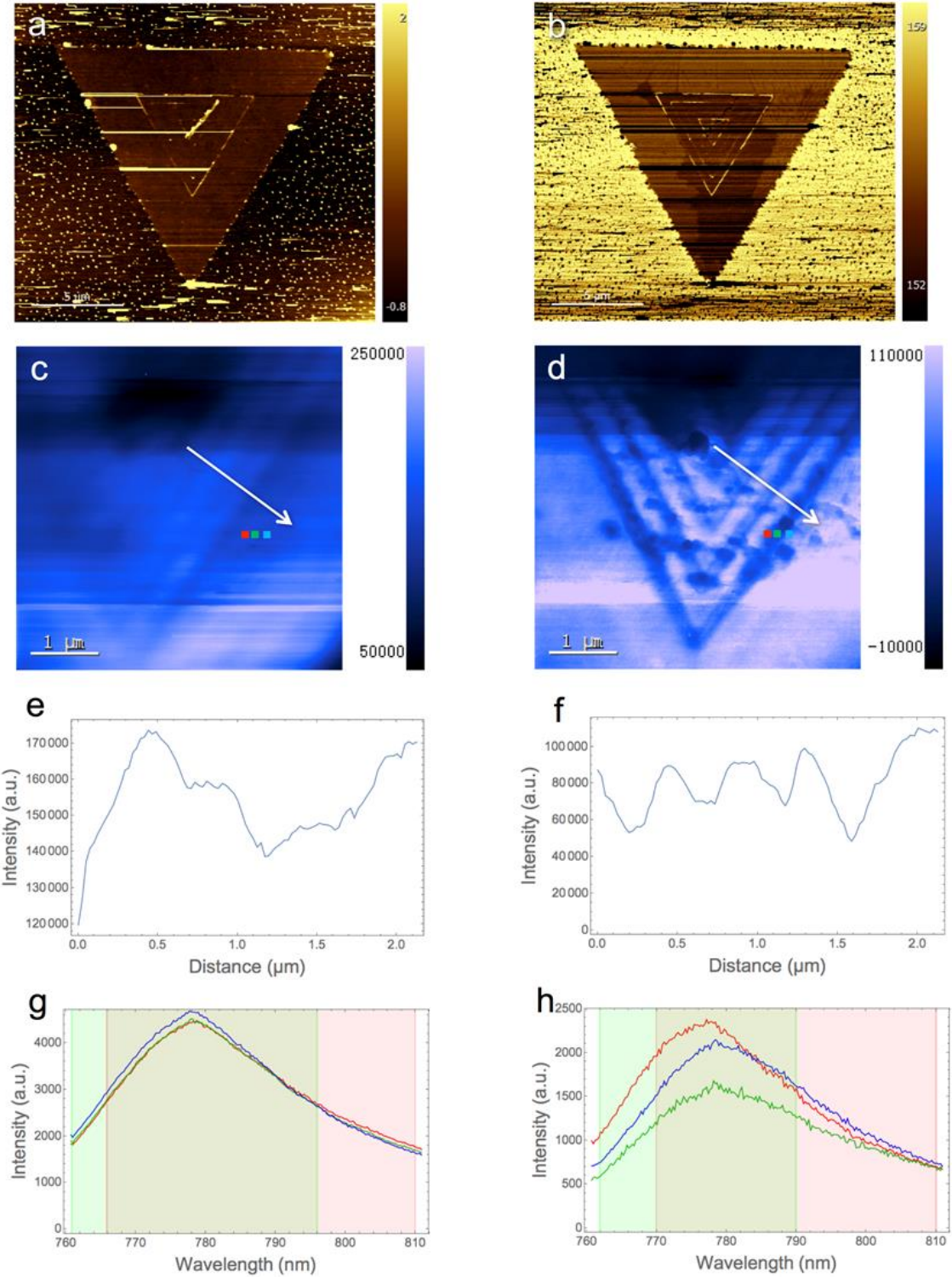

Figure 2. (a) AFM topography map and (b) the corresponding AFM phase map of a monolayer MoSe$_2$-WSe$_2$ lateral heterostructure triangular domain grown by chemical vapor deposition. A partial area was selected to perform TEPL imaging shown in (c) and (d). (c) FF PL intensity map. (d) TEPL NF intensity map. (e) and (f) are line profiles from the areas in (c) and (d) marked by white arrows showing PL intensity variations. (g) and (h) show FF and NF PL spectra of three different points marked by the colored squares in (c) and (d), respectively.

The FF PL spectra recorded from different positions of the lateral heterostructure near the junction do not vary significantly (Fig. 2g), since the focal spot size used for the PL signal acquisition was greater than the average width of the interface. However, the NF PL spectra showed larger variations compared to the FF PL spectra, and exhibit a clear red shift near the heterojunction (Fig. 2h). The dominating emission peak observed at ~ 775 nm corresponds to the emission from a direct exciton in the monolayer $WSe_2$. However, the PL peak corresponding to the $MoSe_2$ domain (~ 810 nm) was weaker than that of the PL of $WSe_2$. The observed red shift at the interface may be due to the alloying between W and Mo around the vicinity of the hetero-interface. Furthermore, the optical transitions at heterojunctions in the NF line profiles are sharper than those in the FF maps. The spatial separation/width of individual TMD domains in the heterostructure is equal to the distance between the two dips in the line profile. Hence, TEPL NF map has higher spatial resolution, and can be used for better understanding of the nanoscale emission characteristics of TMD lateral heterostructures.

By considering different PL spectral ranges, we were able to map the characteristic PL signals originating from different crystalline domains of the $MoSe_2$-$WSe_2$ lateral heterostructure. By the appropriate peak filtering process, these images also showed peak shift maps from different regions of the triangular TMD heterostructure. The nano-PL emission map in the wavelength range of ~ 760 – 790 nm was mostly attributed to the monolayer $WSe_2$ (Fig. 3a). The presence of the sequential $WSe_2$ strips was more prominently resolved. In contrast, the absence of the PL in this spectral range corresponding to $MoSe_2$ domains (black strips in Fig. 3a) further supported the presence of distinct material domains. In contrast, by selecting the range between ~ 770 – 810 nm, the distribution of the nano-PL signals was mostly from the $MoSe_2$ domains (red strips in Fig. 3b). The intense PL emission characteristics of monolayer $WSe_2$ as compared to monolayer $MoSe_2$ make the mapping process more difficult to resolve. Similar spectral range selection was applied to obtain the FF PL maps (Figs. 3c and 3d). However, the spatial resolution was lower in the FF maps that in the NF maps, and the junction strips were not clearly distinguishable. Composite PL maps were obtained by overlapping the PL maps of individual $WSe_2$ and $MoSe_2$ regions as shown in the NF and FF maps in Figs. 3e and 3f. Alternating PL bands (green-$WSe_2$ and red-$MoSe_2$ domains) were clearly observed in the NF PL map but not in the FF map.

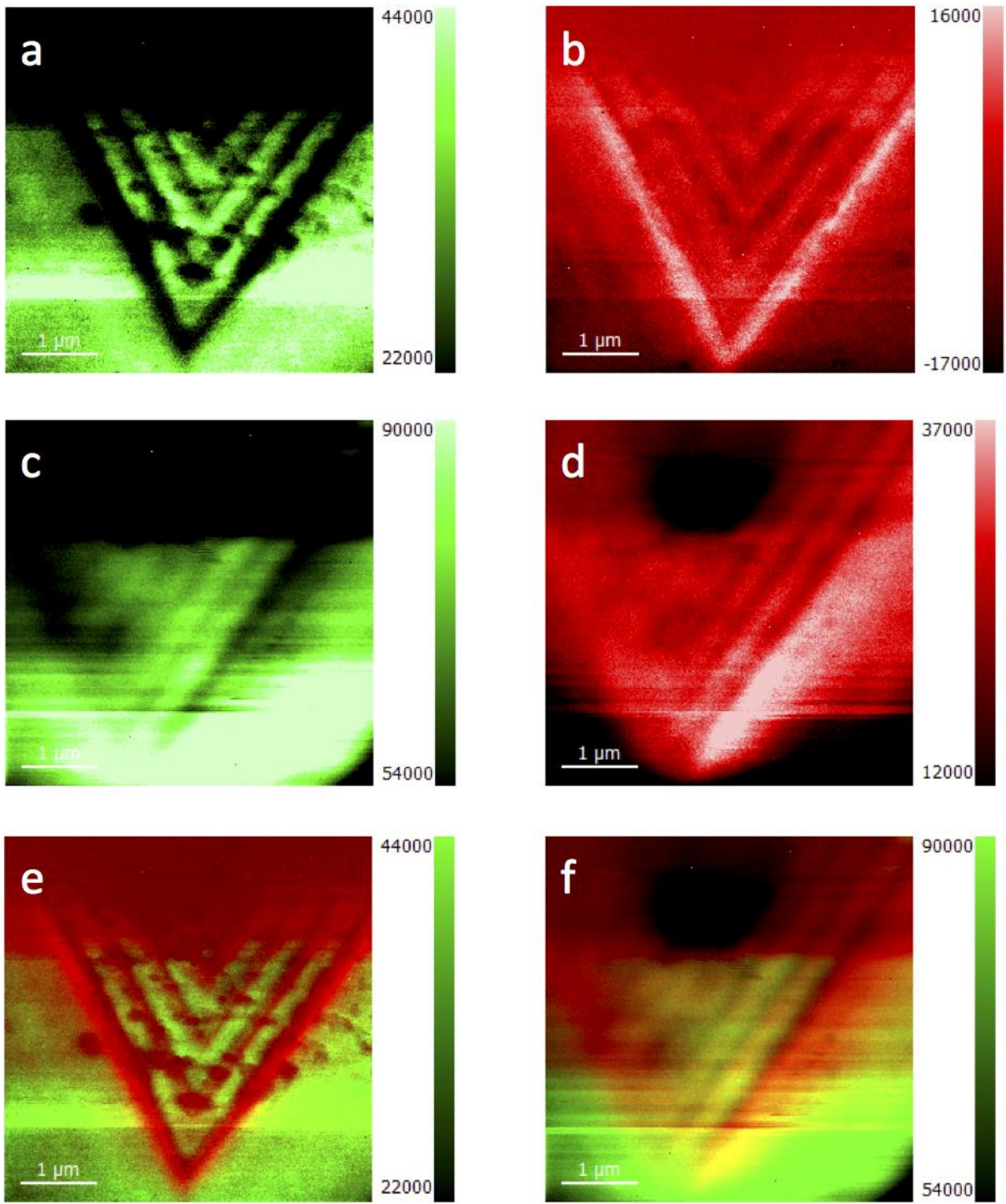

Figure 3. (a) NF PL intensity map in the 762 – 790 nm range (green-monolayer $WSe_2$ domains). (b) NF PL intensity map in the 770 – 810 nm range (red-monolayer $MoSe_2$ domains). (c) FF PL intensity map in the 761 – 796 nm range. (d) FF PL intensity map in the 766 – 810 nm range. (e) Overlap of the NF PL maps in (a) and (b) (green-$WSe_2$ and red-$MoSe_2$ domains). (f) Overlap of the FF PL maps in (c) and (d).

To better visualize the PL spatial distribution, peak positions were mapped both in the FF (Fig. 4a) and NF (Fig. 4b) modes. The peak position shifts in accordance with the change of the material domains. This change can be seen more clearly in the NF map as compared to the FF. A line profile further depicts the change in the peak position across the material domain (Fig. 4c). This change in the nano-PL peak position is slightly different at different heterojunction. It further corroborates with the nature of the crystalline interface i.e., the sharp peak shift is observed while changing the material domain from $WSe_2 \rightarrow MoSe_2$, whereas the observed gradual decrease in the peak shift is referred to material transition from $MoSe_2 \rightarrow WSe_2$ domains. This nature of the interface characteristics depends on the carrier gas switching cycle during the sequential growth process of the TMDs lateral heterostructure.

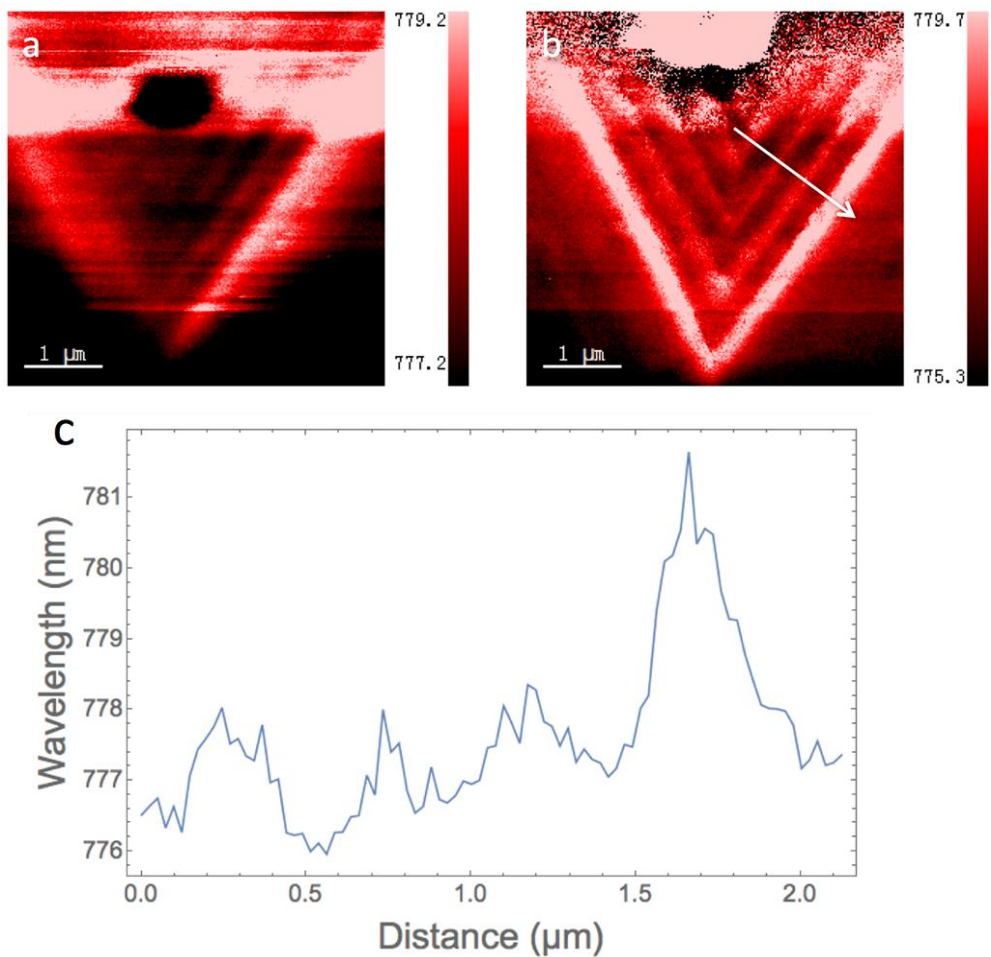

Figure 4. (a) FF and (b) NF peak position maps. Color bar represents PL wavelength in nm. (c) NF line profile showing the variation of the nano-PL peak position across a line in the NF map as indicated by the arrow in (b).

**Conclusions**

Our results provide a new insight into the nanoscale optical properties with sub-diffraction resolution of multi-junction monolayer $MoSe_2$-$WSe_2$ lateral heterostructure grown by CVD. Significant nanoscale heterogeneity in the nano-PL spectra is observed across the interface. NF imaging provides a better spatial resolution compared to FF. Thus, NF nano-PL maps can be used to probe the extent of the material diffusion across the lateral domain of any 2D heterostructure. The variations in the nanoscale optical properties corroborating with the nanoscale structural variation within the lateral heterostructure can provide a better understanding of these materials for the future development of highly efficient optoelectronic devices.

**Acknowledgements**

We thank Marlan Scully, Alexei Sokolov and Zhe He for helpful discussions. D.V.V. acknowledges the support by the National Science Foundation (grant number CHE-1609608). H.R.G. acknowledges support by the National Science Foundation Grant DMR-1557434.